\documentclass{fmj2010}
\usepackage{graphicx}

\def\fermilat{\textit{Fermi}/LAT}
\def\fermi{\textit{Fermi}}

\begin{document}
   \title{Recent multi-wavelength campaigns in the \fermi-GST era}

   \author{Lars Fuhrmann\inst{1} on behalf of the \fermilat\ collaboration \and many
     MW collaborators\thanks{The work reviewed here was possible only
       due to the large efforts and great work of many
       multi-wavelength (MW) projects/collaborators, it is 
       impossible to mention all in detail here. Representatively, we give the
       names of the corresponding projects in the text. The collaborators are part
       of the referenced publications for each campaign.}  }

   \institute{Max-Planck-Institut f\"ur Radioastronomie, Auf dem H\"ugel 69, 53121 Bonn, Germany
             }
             \abstract{Since 2008 the \fermilat\ instrument has delivered highly time-resolved $\gamma$-ray 
	     spectra and detailed variability curves for a steadily increasing number of AGN. For detailed 
	     AGN studies the \fermilat\ data have to be combined with, and accompanied by, dedicated ground- and 
	     space-based multi-frequency observations. In this framework, the \fermi\ AGN team has realized  
             a detailed plan for multi-wavelength campaigns including a large suite of cm/mm/sub-mm band
	     instruments. Many of those campaigns have been triggered, often for sources detected in flaring 
	     states. We review here a few interesting results recently obtained during three such campaigns, 
	     namely for the flat-spectrum radio quasar 3C\,279, the Narrow Line Seyfert 1 PMN\,J0948+0022 
	     and quasar 3C\,454.3.  
             }
	     \authorrunning{L. Fuhrmann}
   \maketitle
%

\section{Introduction}
The successful launch in 2008 and subsequent smooth operation of the \fermi\  
Gamma-ray Space Telescope (\fermi-GST) has brought the community a powerful instrument   
which is monitoring the entire $\gamma$-ray sky about every 3 hours. Thus, as an ``all-sky-monitor", 
\fermilat\ delivers highly time-resolved $\gamma$-ray spectra and detailed variability curves 
for a steadily increasing number of AGN. For the first time, detailed studies of   
AGN properties at $\gamma$-ray energies become possible and already many interesting results 
have been obtained (e.g., Abdo et al. 2009c, Abdo et al. 2010a, Abdo et al. 2010c). However, only 
when combined with, and accompanied by, dedicated ground- and space-based multi-frequency 
observations, can the \fermilat\ unfold its full capability of providing a tremendous opportunity 
for systematic and detailed studies of the physical processes at work in AGN. Consequently, 
a large suite of different multi-wavelength (MW) monitoring data and projects (``single-dish", 
VLBI, polarization, spectra) across the whole electromagnetic spectrum (cm/mm/sub-mm, IR/optical/UV, X-ray, TeV) are essential to complement 
the \fermi\ $\gamma$-ray observations. Together, important fundamental questions about e.g., 
$\gamma$-ray production, overall emission and variability processes as well as the location of the 
$\gamma$-ray emission region can be effectively addressed.  

In this framework, the \fermi\ AGN group has realized a detailed plan for ad-hoc as well as 
intensive long-term campaigns. Many of these have been triggered, often for sources 
detected in flaring states. Here, the 15\,GHz OVRO 40-m and F-GAMMA cm to sub-mm (Effelsberg 100-m, IRAM 30-m, APEX 12-m) monitoring 
programs, the GASP (radio/IR/optical) collaboration including many IR/optical telescopes (radio: UMRAO, Mets\"ahovi, 
SMA, Medicina, Noto), RATAN-600, ATCA, Kanata, ATOM, SMARTS, Stewart observatory, MDM, WIRO, KVA, INAOEP, VLT/VISIR as well 
as VLBI: MOJAVE, TANAMI, the Boston 43\,GHz program, a VLBA multi-frequency ToO program and the EVN/LBA have been 
participating in one or more of the various campaigns. 

In addition the X-ray bands have often been covered by the space-based X-ray observatories 
\textit{Swift}, \textit{Suzaku}, and RXTE. In particular, \textit{Swift} has proven to be extremely valuable in quickly providing 
detailed and simultaneous observations at optical/UV and X-ray bands for many sources. 
Furthermore, \textit{Spitzer} has participated in the case of 3C\,454.3 with important near-IR data. 
Finally, first combined \fermilat\ and TeV campaigns led to joint studies with the Cherenkov 
telescopes HESS and VERITAS as e.g., in the case of PKS\,2155-304 (Aharonian et al. 2009) and 
3C\,66A (Abdo et al. in prep.).

Since the launch of \fermi-GST in 2008, many sources have been target of detailed MW campaigns 
triggered by the \fermi\ AGN group. Table 1 provides a short summary of those which have been 
published so far. Many other MW campaign publications are accepted, have been submitted or 
are in progress, e.g., for the galactic plane source J\,0109+6134 (Abdo et al. 2010d), PKS\,1510-089, Mrk\,501, Mrk\,421
and 3C\,454.3.       
   
As examples, we review here a few interesting results from three selected MW campaigns recently conducted (2008--2009)
by the \fermi\ AGN group together with many MW collaborators.


   \begin{table}
      \caption[]{Publication summary of sources studied by the \fermi~AGN group including 
      (quasi-) simultaneous MW data. Joint GeV/TeV projects are also included.}
         \label{table1}
\centering
\[
\resizebox{\columnwidth}{!}{%
\begin{tabular}[2]{@{}ll@{}}
            \hline
	    \hline
            \noalign{\smallskip}
            Source      &  Reference \\
            \noalign{\smallskip}
	    \hline
            \noalign{\smallskip}
RGB\,J0710+591 & Acciari, V.~A., et~al.\ 2010, \apjl, 715, L49 \\
5 FSRQs        & Abdo,  A.~A., et~al.\ 2010, \apj, 716, 835 \\
PKS 1424+240   & Acciari, V.~A., et~al.\ 2010, \apjl, 708, L100\\
3C\,279        & Abdo, A.~A., et~al.\ 2010, \nat, 463, 919\\
PKS 1502+106   & Abdo, A.~A., et~al.\ 2010, \apj, 710, 810\\
NGC\,1275      & Acciari, V.~A., et~al.\ 2009, \apjl, 706, L275,\\
               & Abdo, A.~A., et~al.\ 2009, \apj, 699, 31\\              
3C\,454.3      & Abdo, A.~A., et~al.\ 2009, \apj, 699, 817\\  
PKS 2155-304   & Aharonian, F., et~al.\ 2009, \apjl, 696, L150\\
PMN J0948+0022 & Abdo, A.~A., et~al.\ 2009, \apj, 707, 727,\\
               & Abdo, A.~A., et~al.\ 2009, \apj, 699, 976\\               
PKS 1454-354   & Abdo, A.~A., et~al.\ 2009, \apj, 697, 934  \\
            \noalign{\smallskip}
            \hline
\end{tabular}
}
\]
   \end{table}


\section{The $\gamma$-ray/optical polarization angle event in 3C\,279}

\begin{figure}[thbp]
   	\includegraphics[clip,width=\columnwidth]{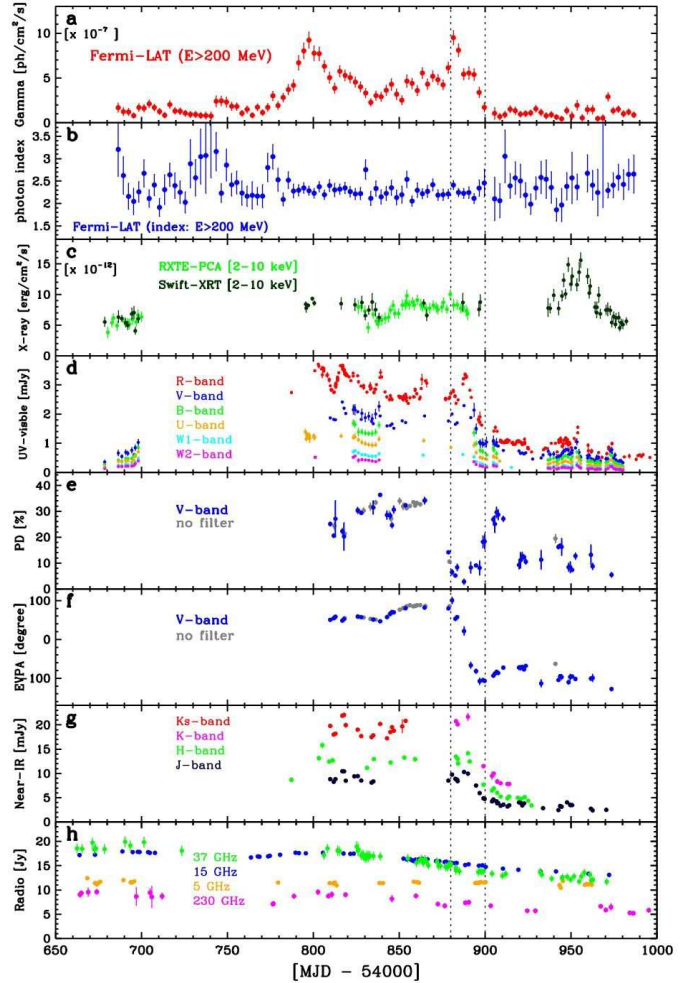}
%
	\caption{
Multi-frequency light curves of 3C\,279 obtained during the large 
		MW campaign between July 2008 and June 2009 (see Abdo et al. 2010b
		for details). Many MW facilities participated such as \textit{Swift}-XRT and 
		RXTE at X-ray bands; many telescopes of the GASP collaboration, Kanata, 
		\textit{Swift} UVOT and KVA at IR/optical/UV bands, as well as SMA, UMRAO, 
		OVRO, Mets\"ahovi, Medicina, Noto and Effelsberg at radio bands.
		Note the smooth optical polarization angle swing 
		during the period of the second, rapid $\gamma$-ray flare (dotted lines).}
            \label{3c279}
    \end{figure}

After about 100 days of \fermilat\ routine operations, the 
quiescent phase of flat-spectrum radio quasar (FSRQ) 3C\,279 
turned into a phase of strong $\gamma$-ray activity and a MW 
campaign was triggered including a large number of instruments
(see Fig.~\ref{3c279}, Abdo et al. 2010b). 

\begin{figure*}[t]
   	\includegraphics[clip,width=0.80\linewidth]{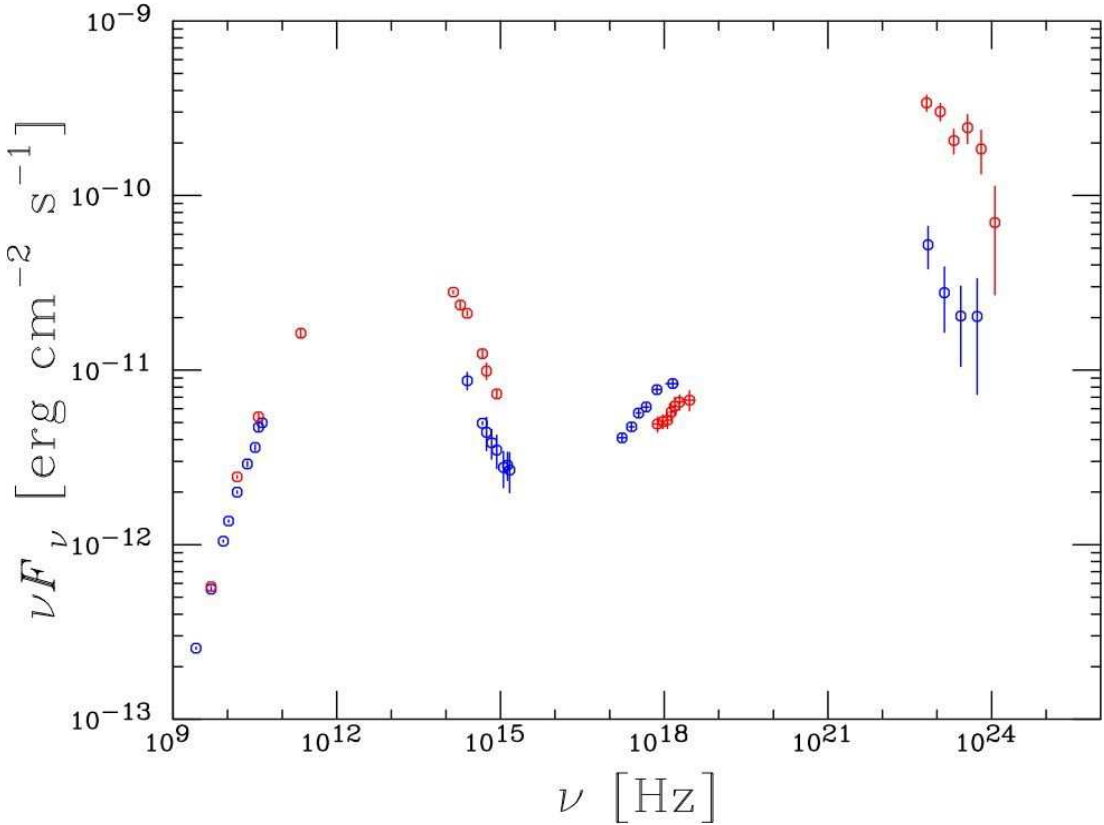}
\hfill\parbox[b]{0.17\textwidth}{\caption[]{
%
	The red data points denote the period of the $\gamma$-ray/optical event. The blue data points 
	have been taken during the period of the isolated X-ray flare (see Abdo et al. 2010b
		for details).}
}
            \label{3c279_SED}
    \end{figure*}

As seen from Fig.~\ref{3c279}, 3C\,279 went into a high $\gamma$-ray
state at around MJD 54780 lasting for about 120 days and characterized
by a double-peak structure with overall variations of the flux by a factor
$\sim$\,10. The observed $\gamma$-ray luminosity of
$\sim\,10^{48}$\,erg\,s$^{-1}$ dominates the power emitted across the
whole electromagnetic spectrum (see Fig.~\ref{3c279_SED}).

%
%
%

%
%
%

The most striking event occured during the rapid, second $\gamma$-ray 
flare (around MJD 54880, doubling time scale of about one day). Here,
a highly correlated behavior of the $\gamma$-ray and optical bands is
evident between MJD 54800 and 54830, with the $\gamma$-ray flare
coincident with a significant drop of the level of optical
polarization, from about 30\,\% down to a few percent lasting for
about 20 days.  In particular, this event is associated with a
dramatic change of the electric vector position angle (EVPA) by
208$^{\circ}$ (12$^{\circ}$/day), in contrast to being relatively
constant earlier, at about 50$^{\circ}$, which corresponds to the jet
direction of 3C\,279 as observed by VLBI. The close association of the
$\gamma$-ray flare with the optical polarization angle event clearly
suggests that the $\gamma$-ray emission was produced in a single,
coherent event and happened co-spatial with the optical. It
furthermore suggests highly ordered magnetic fields in the
$\gamma$-ray emission region.

Compared to the higher energy emission, the radio cm/mm bands showed
less strong variability and no obvious ``correlated event" is evident
from the light curves shown in Fig.~\ref{3c279}. Still, the source
appears to be at higher radio flux levels (factor $\sim$\,2) during
the period of the overall $\gamma$-ray high state as seen from the
230\,GHz SMA data. However, assuming the source was still optically
thick at these bands, synchrotron self-absorption arguments constrain
the transverse size of the emission region to
$<\,5\,\times\,10^{16}$\,cm, in good agreement with the values
obtained from the shortest $\gamma$-ray variability.
      
The gradual rotation of the optical polarization angle requires a
non-axisymmetric trajectory of the emission pattern, since in a uniform,
axially-symmetric case, any e.g., compression due to a shock moving
along the jet would result in a change of polarization degree, but
not in a gradual change of the EVPA. Consequently, two models have
been discussed to explain the observed behavior in a
non-axisymmetric/curved geometry: the emission region/knot propagating
outwards along (i) helical magnetic field lines (similar to the 
optical polarization event observed in BL\,Lacertae, Marscher et al. 
2008) and (ii) along the curved trajectory of a bent jet.

In both scenarios the distance of the dissipation region from the 
central engine can be constrained from the $\sim$\,20\,day time-scale 
of the event. The distance obtained is about 5 orders of magnitude 
larger than the gravitational radius of the black hole in 3C\,279  
and implies a jet opening angle of $<$\,0.2$^{\circ}$, smaller than 
typically observed with VLBI. Although less likely, models resulting 
in a much smaller distance (sub-parsec) can not be completely ruled 
out. At the large distances implied by the two models (parsecs), 
the seed photons for the IC emission should then mostly be provided by the torus 
IR and jet synchrotron emission rather than BLR or accretion disk emission.   

Another interesting feature is the isolated X-ray flare at MJD 54950,
about 60\,days after the second $\gamma$-ray flare. The hard X-ray spectrum 
during this period and the similarity of its shape and time-scale to  
the $\gamma$-ray flare argue in favor of an isolated event which 
is difficult to reconcile with simple one-zone models.

\section{PMN\,J0948+0022 and Narrow-line Seyfert 1 galaxies}

Before the launch of \fermi-GST the known types of $\gamma$-ray
emitting AGN were blazars and radio galaxies. Indeed, the early
\fermilat\ three month results (Abdo et al. 2009c) confirmed that the
extragalactic $\gamma$-ray sky is dominated by radio-loud AGN, being 
mostly blazars and a few radio galaxies. However, an important and
impressive early discovery of {\it Fermi}-GST is the detection of
$\gamma$-rays from a different class of AGN: Narrow Line Seyfert\,1
galaxies (NLS1). These types of objects are believed to be active nuclei
similar to those of Seyferts with optical spectra showing permitted
lines from the broad-line region, although much narrower than
typically seen in Seyfert 1s or blazars
(FWHM(H$\beta)\,<$\,2000\,km\,s$^{-1}$). This and other
characteristics make them a unique class of AGN, whereas a large
fraction is radio-quiet, and only less than 7\,\% (Komossa et
al. 2006) are found to be radio-loud. The first \fermilat\ detection
of $\gamma$-rays from a NLS1, namely in PMN\,J0948+0022 (Abdo et al.
2009b), certainly once more raised the question whether relativistic 
jets exist in this type of object, as indicated by previous studies 
in particular for the most radio-loud NLS1 (e.g., Foschini et al. 2009a).

   \begin{figure*}[t]
   \centering
   \includegraphics[clip,width=0.80\linewidth]{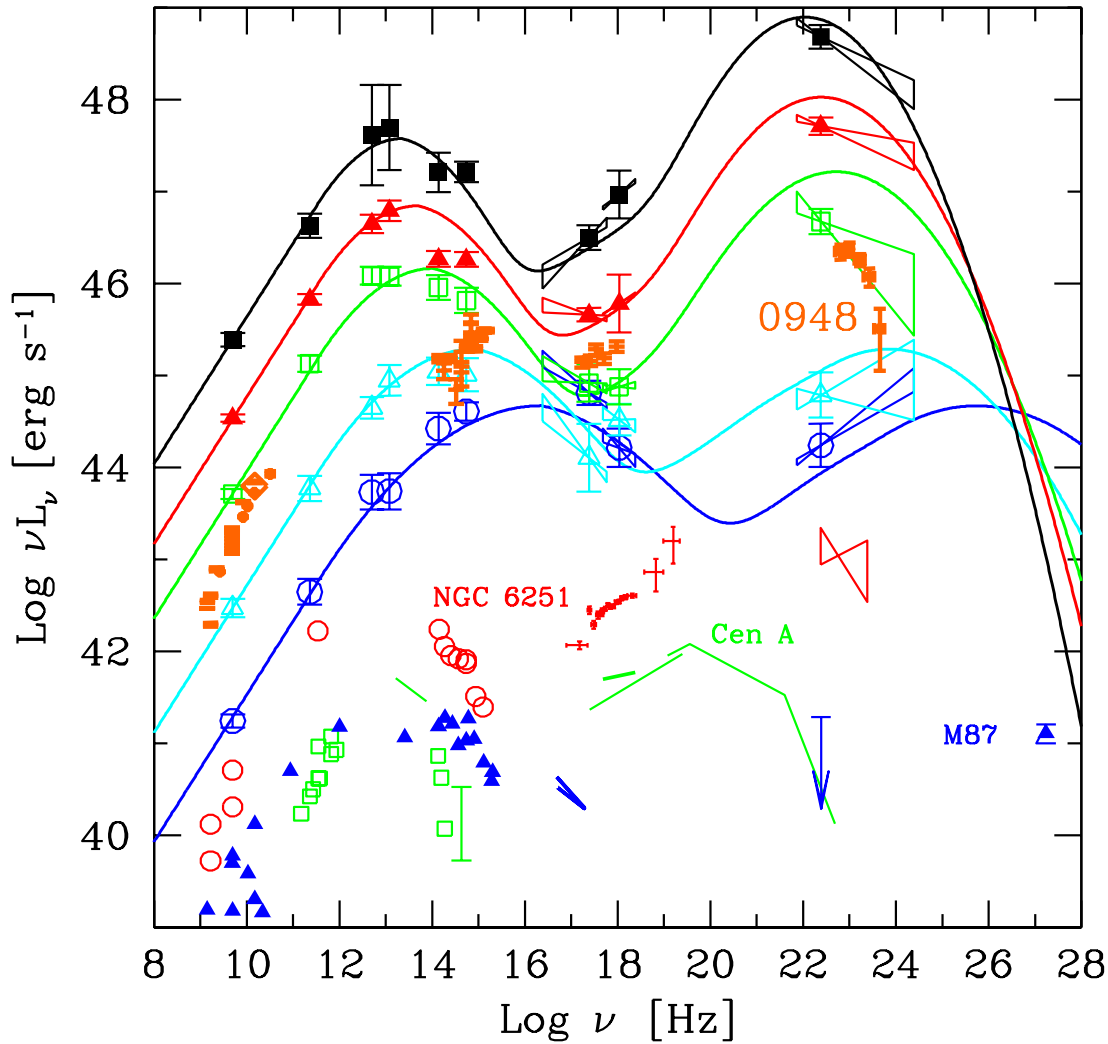}
\hfill\parbox[b]{0.17\textwidth}{\caption[]{
The SED of PMN\,J0948+0022 as obtained during the MW campaign, here shown in 
      	       comparison to other $\gamma$-ray emitting FSRQs, BL\,Lacs and radio galaxies
	       (from Foschini et al. 2009b, see also Abdo et al. 2009b, Abdo et al. 2009d).}
}
         \label{0948_SED}
   \end{figure*}
  
The answer came promptly. MW follow-up observations of PMN\,J0948+0022
performed right after its $\gamma$-ray detection (Abdo et al. 2009b) as 
well as through a triggered MW campaign during March--July 2009 (Abdo et
al. 2009d) demonstrated the efficiency of MW observations/campaigns in
conjunction with \fermilat: these MW studies have demonstrated 
that PMN\,J0948+0022 hosts a relativistic jet. Here, early SED studies 
using non-simultaneous plus simultaneously acquired MW data (Effelsberg 
100-m, OVRO 40-m, \textit{Swift} satellite) revealed an SED similar to 
that of powerful FSRQs with the typical double-humped appearance peaking in the 
far-IR and in the 40--400\,MeV range (see Fig.~\ref{0948_SED}). Signs 
of the accretion disk peaking in the \textit{Swift} UV frequency range are clearly 
seen, which yields a lower limit to the black hole mass of 
$1.5\,\times\,10^{8}$\,M$_{\odot}$. The time-resolved SEDs have been fitted 
using the one-zone synchrotron/IC model of Ghisellini \& Tavecchio 
(2009) resulting in synchrotron/SSC components (dominating the radio 
to X-ray frequencies) and an EC component producing the $\gamma$-ray 
emission. The physical parameters are similar to those of blazars, 
however, with lower power compared to FSRQs but higher values than 
typically seen for BL\,Lacs (see Fig.~\ref{0948_SED}).     
   
From the radio perspective alone, the presence of a relativistic jet
in PMN\,J0948+0022 appears obvious due to several findings, such as
(i) flux density as well as spectral variability/flare over the duration 
of the campaign with flat ($\alpha_{5-15\,\mathrm{GHz}}\sim 0$) to highly inverted
(max.:\,$\alpha_{5-15\,GHz}=0.98\pm0.05$) Effelsberg/RATAN radio spectra, 
(ii) equipartition Doppler factors of up to $\sim$\,7, (iii) a highly
compact, unresolved core on pc-scales (MOJAVE VLBA and EVN/LBA) with 
a 15\,GHz core size of $<$\,60\,$\mu$as and corresponding core brightness 
temperature of $1.0\times 10^{12}$\,K and finally, (iv) VLBI core 
fractional linear polarization of 0.7\%. This is the signature of  
a relativistic radio jet similar to those seen in powerful blazar type objects. 
The radio flare seen in the OVRO/Mets\"ahovi light curves as well as 
Effelsberg/RATAN radio spectra appears to be delayed with respect 
to the $\gamma$-ray peak by 1.5--2 months.     

In summary, the \fermilat\ and MW observations of PMN\,J0948+0022
clearly demonstrate for the first time the existence of a $\gamma$-ray 
emitting NLS1 hosting a relativistic jet similar to blazars even though 
the environment in the vicinity of the central engine is most likely 
pretty different. However, this strongly challenges our view that jets 
can only develop in elliptical galaxies. Follow-up \fermilat\ and MW 
observations of PMN\,J0948+0022 and the three other NLS1 detected 
by \fermilat\ (Abdo et al. 2009e) will certainly shed further light 
on this interesting new type of $\gamma$-ray emitting AGN.  

\section{The early $\gamma$-ray flare of 3C\,454.3 during 2008}

\begin{figure}
\centering
\includegraphics[clip,width=\columnwidth,angle=0]{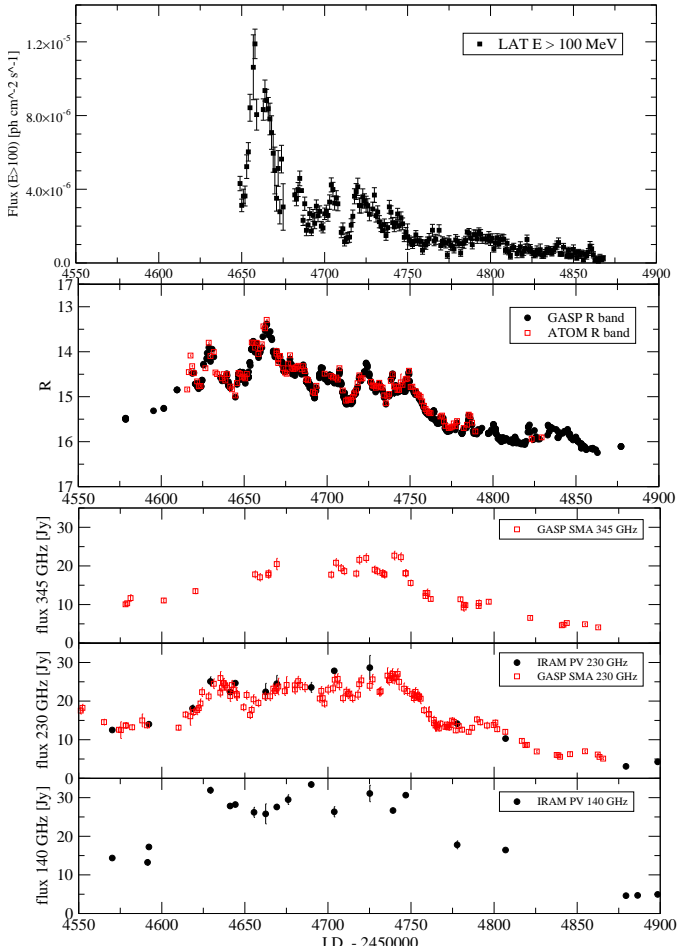} 
\caption{Selection of multi-band light curves for 3C\,454.3 (Abdo et al. in prep.)
obtained at $\gamma$-rays, optical R band (GASP and ATOM telescopes) and mm/sub-mm bands (SMA, IRAM 30-m).
Note the similar variability pattern.}
\label{3c454_lc1}
\end{figure}

During the early check-out phase of \fermilat\ and the subsequent
early operation in survey mode (July-October 2008), strong and highly
variable $\gamma$-ray emission from the quasar 3C\,454.3 was detected
(Abdo et al. 2009a) showing a large outburst in July 2008 and
subsequently, distinct symmetrically shaped sub-flares on time scales
of days (see Fig.~\ref{3c454_lc1}).

Such rapid $\gamma$-ray variability indicates a highly compact 
emission region and relativistic beaming with a Doppler factor of
$\delta\,>\,8$ in order to be optically thin to pair production.
The observed $\gamma$-ray spectrum obtained from the early \fermilat\ 
data has demonstrated for the first time the existence of a spectral 
break for a high-luminosity blazar above 100\,MeV, which may be regarded 
as evidence for an intrinsic break in the energy distribution of 
the radiating particles. Alternatively, the spectral softening above 
2\,GeV could be due to $\gamma$-ray absorption via photon-photon pair 
production on the soft X-ray photon field of the host active galactic 
nucleus or due to the superposition of two different spectral 
components at high energies (Abdo et al. 2009a, see also Finke \& Dermer 
2010 for a more detailed study of the spectral break).

The large multi-wavelength campaign (Abdo et al. in prep.) triggered by 
the Fermi AGN team shortly after the detection of the high $\gamma$-ray 
state of the source in July/August 2008, resulted in a so far unprecedented 
frequency coverage from cm/mm/sub-mm bands, IR/optical/UV, X-ray to the GeV range. 
These data sets demonstrate an active phase of 3C\,454.3 across the
whole electromagnetic spectrum. Figure~\ref{3c454_lc1} shows a selection 
of MW light curves including the $\gamma$-ray, optical (R) and 
short-mm bands. Interestingly, a similar variability pattern is 
seen at all bands, even down to the short-mm bands - up to frequencies 
higher than or close to the radio synchrotron turn-over at around 100\,GHz.     
A detailed time series and cross-band analysis of the best sampled 
light curves reveals (i) strong correlations between $\gamma$-ray/optical 
and optical/mm-bands (ii) a quasi-periodic modulation of the variability with a 
fast (about 20\,days) and slow (about 60\,days) component seen at 
all bands, with similar start and stop times. The optical polarization 
angle data, however, shows no obvious strong pattern, in contrast to 
the case of 3C\,279.  

   \begin{figure*}
   \centering
   \includegraphics[trim=50 10 65 5, clip, width=0.80\linewidth]{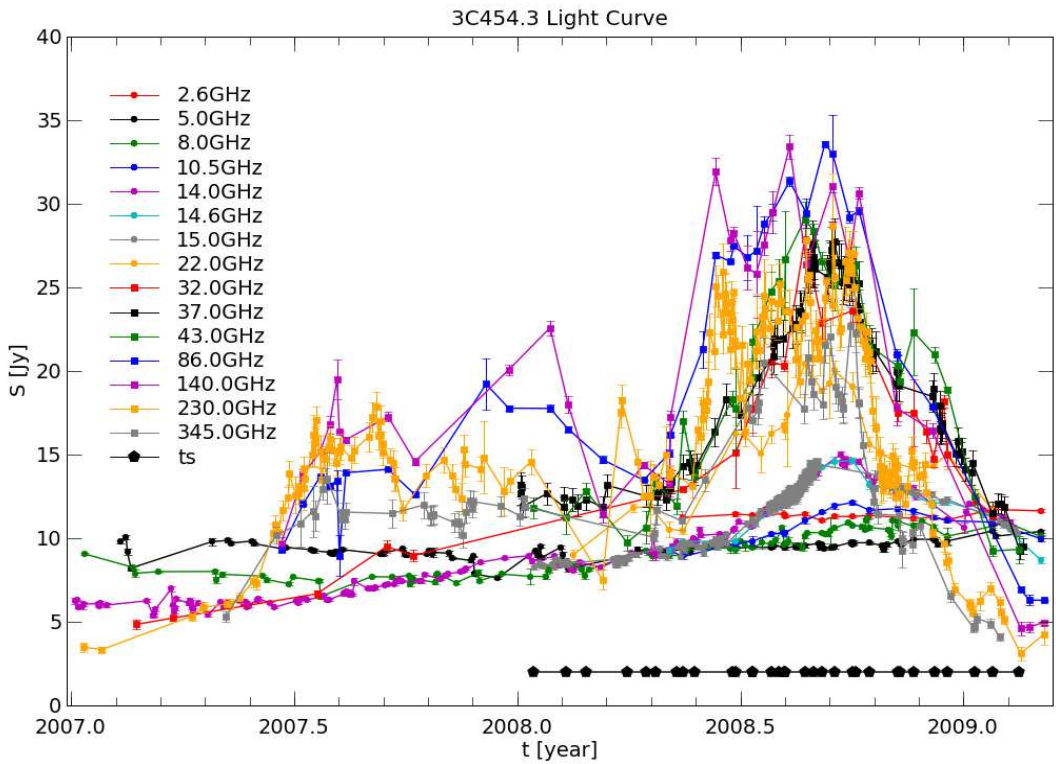}
\hfill\parbox[b]{0.17\textwidth}{\caption[]{
      The full collection of multi-frequency radio cm/mm/sub-mm light curves 
      of 3C\,454.3 obtained during the MW campaign (Abdo et al. in prep.). Many 
      radio telescopes were involved:
      UMRAO, Medicina, Noto, Effelsberg, OVRO, Mets\"ahovi, IRAM 30-m, SMA. The great frequency 
      coverage has been used to study detailed spectral evolution.}
}
         \label{3c454_lc2}
   \end{figure*}

Doppler factors in the range of 3--9 derived from the radio variability 
(Fig.~\ref{3c454_lc2}) are in good agreement with those obtained from synchrotron 
self-absorption and $\gamma$-pair production arguments. Three epochs of 
multi-frequency VLBA ToO observations clearly show that the total single-dish 
variability originates from the core region while the core spectrum nicely resembles 
the (inverted) total single-dish spectrum.       

The evolution of the synchrotron turnover frequency as obtained from the detailed 
radio light curves shown in Fig.~\ref{3c454_lc2} is in good agreement with the 
shock-in-jet model of Marscher \& Gear (1985) (as are the increasing time lags 
towards longer radio wavelengths), at least in the synchrotron and adiabatic phases. 
However, departures from the Compton phase indicate additional processes at work 
as indicated already by the cross-band analysis. Detailed modeling with geometrical (e.g., 
helical jet) as well as SSC/EC models are in progress in order to explain the 
complex behavior of the source in a consistent manner.

\section{Conclusions}
As a powerful ``all-sky-monitor" the \fermilat\ instrument provides a unique 
opportunity to explore the high energy $\gamma$-ray sky and the $\gamma$-ray 
characteristics of the AGN population. In particular, when combined with 
ground- and space-based multi-frequency observations, \fermilat\  unfolds its 
full capability in addressing fundamental questions about energy production 
in AGN. This becomes possible due to the large efforts of the MW community in 
providing detailed, (quasi-) simultaneous broad-band data for a large number 
of \fermi-detected AGN. Since 2008 the \fermi\ team
has triggered a large number of 
MW campaigns. The success of such campaigns, although challenging for both 
observers and 
theoreticians---as demonstrated by the examples presented 
here---increasingly provides deeper insight into the physical processes involved.    

\begin{acknowledgements}
The Fermi/LAT Collaboration acknowledges support from
a number of agencies and institutes for both development
and the operation of the LAT as well as scientific data 
analysis. These include NASA and DOE in the United States,
CEA/Irfu and IN2P3/CNRS in France, ASI and INFN in Italy,
MEXT, KEK, and JAXA in Japan, and the K. A. Wallenberg
Foundation, the Swedish Research Council and the National
Space Board in Sweden. Additional support from INAF in Italy
and CNES in France for science analysis during the operations
phase is also gratefully acknowledged.
\end{acknowledgements}

\end{document}